\documentclass[conference]{IEEEtran}
\IEEEoverridecommandlockouts

\usepackage{cite}
\usepackage{amsmath,amssymb,amsfonts}
\usepackage{algorithmic}
\usepackage{graphicx}
\usepackage{textcomp}
\usepackage{xcolor}
\usepackage{booktabs}
\usepackage{tabularx}
\usepackage{pifont} 
\usepackage[table]{xcolor} 
\usepackage{flushend}

\newcommand{\yes}{\ding{55} }
\newcommand{\no}{}

\definecolor{lightgray}{gray}{0.95}

\usepackage{mysty}
\def\BibTeX{{\rm B\kern-.05em{\sc i\kern-.025em b}\kern-.08em
    T\kern-.1667em\lower.7ex\hbox{E}\kern-.125emX}}
\begin{document}


\title{A Comparative Study of MAP and LMMSE Estimators for Blind Inverse Problems  \\

\thanks{This work was supported by the funding received from the European Research Council (ERC) Starting project MALIN under the European Union’s Horizon Europe programme
(grant agreement No. 101117133). The European Commission and the other organizations are not responsible for any
use that may be made of the information it contains.}
}

\author{\IEEEauthorblockN{Nathan Buskulic}
\IEEEauthorblockA{MaLGa, Centre, \\\textit{DIBRIS, Università di Genova}\\
Genova, Italy \\
nathan.buskulic@proton.me}
\and
\IEEEauthorblockN{Luca Calatroni}
\IEEEauthorblockA{MaLGa, Centre, \textit{DIBRIS, Università di Genova}\\
\IEEEauthorblockA{MMS, Italian Institute of Technology}
Genova, Italy \\
luca.calatroni@unige.it
}}

\maketitle

\begin{abstract}
Maximum-a-posteriori (MAP) approaches are an effective framework for inverse problems with known forward operators, particularly when combined with expressive priors and careful parameter selection. In blind settings, however, their use becomes significantly less stable due to the inherent non-convexity of the problem and the potential non-identifiability of the solutions. (Linear) minimum mean square error (MMSE) estimators provide a compelling alternative that can circumvent these limitations.
In this work, we study synthetic two-dimensional blind deconvolution problems under fully controlled conditions, with complete prior knowledge of both the signal and kernel distributions. We compare tailored MAP algorithms with simple LMMSE estimators whose functional form is closely related to that of an optimal Tikhonov estimator. Our results show that, even in these highly controlled settings, MAP methods remain unstable and require extensive parameter tuning, whereas the LMMSE estimator yields a robust and reliable baseline. Moreover, we demonstrate empirically that the LMMSE solution can serve as an effective initialization for MAP approaches, improving their performance and reducing sensitivity to regularization parameters, thereby opening the door to future theoretical and practical developments.
\end{abstract}

\begin{IEEEkeywords}
Blind inverse problems, MAP estimator, LMMSE estimator
\end{IEEEkeywords}

\section{Introduction}

Inverse problems appear everywhere in science when one whishes to recover an underlying signal $\xvc\in X$ from some noisy observations $\yv\in Y$. The special case of deconvolution that we study here is written more formally as:
\begin{align*}
\yv = \hvc \ast \xvc + \veps
\end{align*}
where $\hvc\in H$ is a convolution kernel and $\veps\in E$ is some additive noise. In the following we will study 2D images, where we consider $X = Y = E =  \R^{n\times n}$. and $H = \R^{d\times d}$. 

The classical (or non-blind) setting assumes full knowledge of both $\yv$ and $\hvc$ and has been extensively studied with many algorithms available and theoretical guarantees~\cite{benning2018modern,arridge_solving_2019}. However, many experimental fields do not have access to the true $\hvc$ and have some uncertainty regarding the kernel. For example, in microscopy, the point-spread function (thus, $\hvc$), will depend on the observed sample as it scatters light uniquely. This is known as the blind case where one has to recover a pair $(\xv, \hv)$, which comes with new difficulties such as ambiguous solutions (an example is the no-blur solution where $(\xvc,\hvc)$ and $(\yv,\delta)$ with $\delta$ a Dirac mass produce the same observation).

If we consider that $\xvc$ and $\hvc$ are realizations of random variables following distributions $p(\xv)$ and $p(\hv)$, then the goal of reconstruction algorithms is to obtain some information about the posterior distribution $p(\xv,\hv\vert\yv)$ which is often retrieving a representative sample. A widely used point estimator is the Maximum-a-Posterior one (MAP) which is given by $\Argmax_{\xv,\hv}p(\xv,\hv\vert\yv)$. In the non-blind case this often leads to state-of-the-art reconstruction~\cite{kamilov2023plug}, but it has been shown to be unstable or unsuited in blind settings\cite{nguyen2025diffusion, benichoux2013fundamental} due to the new non-convex nature of the underlying optimization. 

This calls for the study of other type of estimators in the blind setting. A good candidate is the famous Minimum Mean Square Error (MMSE) which is given by $\Expect{\xv,\hv\vert\yv}$. However, this estimator is often harder to obtain as it requires to integrate over the space $X$ and $H$. A simplified case known as the Linear MMSE (LMMSE) estimator allows for a closed-form formula, recently derived for the blind setting~\cite{buskulic2025sample}, and has been extensively studied in the non-blind case being at the basis of Wiener-Kolmogorov~\cite{wiener1949extrapolation} and Kalman filters (see~\cite{kailath2001linear} for a review over linear estimator theory).


\subsection{Contributions}
 In this work, we compare MAP and LMMSE estimators for blind imaging inverse problems using simulated data, in an idealized setting where the full statistical model is known and controlled. This setting enables the derivation of exact MAP formulations and closed-form LMMSE expressions.

We focus, in particular, on two-dimensional blind image deconvolution problems, and perform both qualitative and quantitative comparisons of MAP and LMMSE approaches with respect to multiple criteria, including reconstruction quality for both $\xv$ and $\hv$, computational cost, and sensitivity to hyperparameter tuning. Notably, we observe that the LMMSE solution provides a robust initialization for the non-convex MAP optimization, helping to alleviate both its instability and its sensitivity to regularization parameters.

\section{MAP \& (L)MMSE estimators for blind inverse problems}

We describe both the MAP and the (L)MMSE approach in the context of blind inverse problems, highlighting both  their advantages and drawbacks. Table~\ref{tab:summary} summarizes these features. 

\subsection{MAP estimators}

In blind scenarios, MAP approaches follow from the use of the Bayes rule $p(\xv,\hv\vert\yv) \propto p(\yv\vert\xv,\hv)p(\xv)p(\hv)$. From there, and upon suitable assumptions on the distributions considered (typically, to belong to the exponential family) one can define tailored energy minimization schemes~\cite{geman1984stochastic}. This leads to variational formulations of the type
\begin{align}\label{eq:MAP}\tag{MAP} 
\min_{\xv,\hv}~ D(\hv \ast \xv, \yv) + \lambda_X R_X(\xv) + \lambda_H R_H(\hv)
\end{align}
where $D$ is a data-fidelity term  corresponding, and $R_X$ and $R_H$ are regularization terms that correspond to the \textit{a-priori} knowledge on both $p(\xv)$ and $p(\hv)$. While the choice of $D$ is mostly driven by the likelihood $p(\yv\vert\xv,\hv)$ determined by the noise, the choice of $R_X$ and $R_H$ which could well represent prior knowledge on the solution varies between approaches, which could be either handcrafted~\cite{benning2018modern} or learned from data~\cite{arridge_solving_2019}. Finally, $\lambda_X\in\R^*_+$ and $\lambda_H\in\R^*_+$ are regularization parameters that balance the strength of the prior distributions with respect to the likelihood. Their choice often depends on available (if any) prior assumptions on the noise level and could be tedious in practical applications.

For non-blind inverse problems, this type of MAP approaches led to state-of-the-art results, and can also, under some convexity assumptions on $D$ and $R_X$, be equipped with global convergence guarantees. However, the blind case is much more challenging in that regard as the addition of the new term $R_H$ (even if convex) makes the problem non-convex~\cite{levin2009understanding}, which is added on top of geometric constraints such that $\hv$ has to lie on the simplex. This poses new optimization challenges to escape bad local minima (such as the no-blur solution), and make the result of the procedure potentially very dependent on the initialization point. Furthermore, the result is also dependent on the choice of the pair $(\lambda_X,\lambda_H)$, which is hard to select a-priori, without a cross-validation strategy on accessible ground-truth data for the problem at hand. Finally, the iterative nature of these energy minimization schemes can make them quite computationally demanding.

Despite these difficulties, MAP approaches are still one of the main approaches in blind deconvolution~\cite{chihaouiBlindImageRestoration2024, li2024blinddiff}, which justifies our interest in experimentally comparing them with an alternative baseline method relying on MMSE estimations, in a controlled setting where we can derive tailored expressions for $D$, $R_X$ and $R_H$. 

\subsection{(L)MMSE estimators}



MMSE estimators have drawn for a long time the attention of the inverse problem community, as they use the total probability mass of the distribution and not only its mode, and represent the optimal estimator in the $L^2$ norm. Nowadays, they are at the heart of the modern diffusion-based generative models heavily used in image reconstruction, as the score of a distribution, which is necessary to build these models, is directly linked to the MMSE denoiser through Tweedie's formula. While these MMSE denoisers are not directly accessible as they require to compute integrals over very high-dimensional spaces, they are approximated sufficiently well by neural networks trained on vast datasets. 

While in general MMSE estimators are intractable, one can derive a simplified MMSE-based estimator where we only consider reconstruction that are linear combinations of $\yv$. These Linear MMSE estimators admit a closed-form formula and were recently adapted to the blind setting in~\cite{buskulic2025sample}. In that setting, vectorized estimations for both the unknown signal and the kernel can be obtained by using the following formulas
\begin{align}\tag{LMMSE}\label{eq:LMMSE}
\begin{split}
\widehat{\xv} &= \Expect{\xv} + \Cxy\Cyy\inv \pa{\yv - \Expect{\yv}}\\
\widehat{\hv} &= \Expect{\hv} + \Chy\Cyy\inv \pa{\yv - \Expect{\yv}},
\end{split}
\end{align}
where $\Cxy$ (resp. $\Chy$) is the cross-covariance matrix between $\xv$ and $\yv$ (resp. $\hv$ and $\yv$), and $\Cyy$ is the covariance matrix of $\yv$. In this work, we consider $\Cyy$ to be invertible, which is the case when the noise covariance matrix is full-rank. Otherwise, one could define suitable regularization strategies based on the use of pseudo-inverse or explicit regularization, see ~\cite{buskulic2025sample}.

Despite its apparent simplicity due to its linearity w.r.t.~$\yv$, the estimators \eqref{eq:LMMSE} play the role here as baselines for comparison with the MAP approaches.  Once sufficient knowledge on the distributions of both $\xv$ and $\hv$ is available, such estimators are indeed very simple to compute; they are linked to the widely used Wiener filters, and can be proven to be the solution of an optimal generalized Tikhonov problem (here only to retrieve $\xv$ but details to retrieve $\hv$ are in~\cite[Lemma 2.2]{buskulic2025sample}):
\begin{align*}
    &\min_{\xv} \norm{\Expect{\hv} \ast \xv - \yv}^2_{\Cv_p\inv} + \norm{\xv - \Expect{\xv}}^2_{\Cxx\inv}\\
    \text{where } &\Cv_p = \Cyy - \Expect{\pa{\Expect{\hv}\ast\xv}\pa{\Expect{\hv}\ast\xv}\tp} + \Expect{\yv}\Expect{\yv}\tp
\end{align*}
where here the convolution has to be understood as returning the vectorized result.
This equip this estimator with a lot of theoretical guarantees w.r.t., e.g., to convergent regularization properties equipped with suitable source conditions~\cite{kailath2001linear}. Furthermore, once the (cross-)covariance matrices are calculated, the computation of $\widehat{\xv}$ and $\widehat{\hv}$ can be performed in closed-form.

\begin{table}[t!hbp]
\centering
\caption{Drawbacks of MAP and LMMSE.}
\label{tab:summary}
\begin{tabular}{lcc}
\toprule
\textbf{Drawbacks} & \textbf{MAP Approaches} & \textbf{LMMSE} \\ \midrule
Linear Estimator   & \no & \yes  \\
Iterative procedure                    & \yes  & \no \\
Expert Choice of Prior        & \yes & \no  \\
Non-convex optimization                & \yes & \no  \\
Requires Prior Covariance Matrices      & \no  & \yes \\
Hyperparam. Tuning ($\lambda$, optimization) & \yes & \no  \\
\bottomrule
\end{tabular}
\end{table}

\section{Experimental setting}

In the following, we focus on an illustrative 2D deconvolution problem with synthetic images and Gaussian convolution kernels, in order to have full control over the distribution of both the unknown image $\xvc$ and kernel $\hvc$. Such knowledge allows us to consider the ideal scenario where, on the one hand, tailored regularization terms can be defined within the MAP setting \eqref{eq:MAP}, while, as far as the LLMSE estimator \eqref{eq:LMMSE} is concerned, all expected values and (cross)-covariance
matrices can be explicitly computed. We stress that this is indeed an ideal scenario, since since such knowledge is typically not available in real applications where sampling strategies for $\xv$, $\hv$ and $\yv$ should be considered instead. 

\subsection{Data generation}
We generate images of size $n\times n$ by considering $K$ vectors at random from the orthonormal basis given by the discrete cosine transform (DCT) which gives us a dictionary matrix $\Dv\in R^{K\times N}$, with $N=n^2$. We then generate a random vector $\boldsymbol{\mu_\alpha}$ where each entry is chosen uniformly in $\{\frac{1}{2}, -\frac{1}{2}\}$. Both $\Dv$ and $\boldsymbol{\mu_\alpha}$ are fixed and assumed known once generated. To generate a reference image $\xv$, we thus first draw a vector $\boldsymbol{\alpha} \in \R^K$ from a Laplace distribution located at $\boldsymbol{\mu_\alpha}$ with scale $b>0$, and set $\xvc = \Dv\tp\boldsymbol{\alpha}$. We consider convolution kernels of reduced size ${15\times 15}$ sampled as discretized isotropic Gaussians, with weights normalized to sum to one, i.e., ${\hvc}_{i,j}
 \propto \exp\pa{-\frac{i^2+j^2}{2\sigma^2}}$ with $(i,j)$ the pixel position. The variance parameter $\sigma$ is drawn i.i.d from a Gamma distribution with shape $a$ and rate $\beta$. Finally, additive white Gaussian noise $\veps$ is drawn from $\mathcal{N}(0,c_{\veps}\Id)$ with $c_{\veps}>0$.

\subsection{Modeling \& Algorithmic setup}

Within this controlled setting, we can thus derive explicit $D$, $R_X$ and $R_H$ terms for the MAP approach corresponding to the true negative log distributions. As $\xv$ is entirely determined by $\boldsymbol{\alpha}$, we will now refer to optimization over $\boldsymbol{\alpha}$ and not $\xv$, with a change from $\lambda_X$ to $\lambda_\alpha$ and from $R_X$ to $R_\alpha$. In particular, we choose $D(\boldsymbol{\alpha}, \hv,\yv)=\norm{\hv\ast \Dv\tp\boldsymbol{\alpha} - \yv}^2$ to adapt to the Gaussianity of the noise and $R_\alpha=\norm{\boldsymbol{\alpha} - \boldsymbol{\mu_\alpha}}_{1}$ due to the Laplace prior over $\boldsymbol{\alpha}$. Regularizing directly over the space of $\boldsymbol{\alpha}$ means that we already know the sparse basis on which $\xv$ can be decomposed which is ideal, and rarely the case in real-world applications. Note also that we do not use the theoretical regularization parameters given by the Bayesian interpretation as we choose to keep our two parameters $\lambda_\alpha$ and $\lambda_H$ free which allow a finer control over the algorithm. Concerning the term $R_H$, we will consider two different version that correspond either to the true negative log prior imposed on the Gaussian spread parameter $\sigma$ or to a smooth term more used in the relevant literature:
 \begin{itemize}
 \item $\text{MAP}_\sigma$ relies on the  Gaussian structure of the blur and optimize directly over its variance parameter $\sigma$. Following the Gamma distribution assumption, there follows that $R_H(\hv) = R_H(\sigma) = \beta\sigma - (a - 1)\log(\sigma)$. 
 \item $\text{MAP}_\hv$ relies on a smooth a-priori on $\hv$ and uses $R_H = \norm{\nabla\hv}^2$, thus enforcing only smoothness without any further structural prior.  
 \end{itemize}
We can now define the MAP optimization problem adapted to our setting as:
\begin{align*}
\Argmin_{\boldsymbol{\alpha}\in\R^K,\hv\in S} \norm{\hv\ast \Dv\tp \boldsymbol{\alpha} - \yv}^2 + \lambda_\alpha \norm{\boldsymbol{\alpha} - \boldsymbol{\mu_\alpha}}_{1} + \lambda_H R_H(\hv),
\end{align*}
with $S$ the simplex and $R_H$ chosen accordingly to $\text{MAP}_\sigma$ or $\text{MAP}_\hv$ .

Both approaches are optimized using an alternating minimization scheme~\cite{sroubek2011robust} where optimization over $\boldsymbol{\alpha}$ for a few steps is followed by one step minimization over the kernel, and the  procedure is repeated until convergence. The optimization over $\boldsymbol{\alpha}$ is done using a proximal gradient descent scheme with explicit proximal step~\cite{combettes2005signal}. As far as the kernel optimization is concerned, in $\text{MAP}_\sigma$, we use gradient descent to optimize over $\sigma$, with the constraint that $\sigma > 0$ (ensuring the simplex constraint), while in $\text{MAP}_\hv$,  gradient descent is combined with a projection over the simplex~\cite{condat2016fast} to guarantee the non-negativity and the sum-to-one condition of the kernel.

\smallskip

LLMSE estimates are calculated by \eqref{eq:LMMSE} using insights from  ~\cite[Section 4]{buskulic2025sample} on how to calculate it efficiently for convolution operators.

\smallskip

Given the intrinsic non-convexity of the MAP problem \eqref{eq:MAP} in the two settings above, initialization plays a prominent role for performance.
To mitigate this effect by exploiting good behaviour of the LLMSE estimator, we will consider the two MAP variants $\text{MAP}_\sigma^{\text{boost}}$ and $\text{MAP}_\hv^{\text{boost}}$ which simply run MAP as above, but initialize the optimization with $\widehat{\xv}$ and $\widehat{\hv}$ given by the (inexpensive) application of the LMMSE approach. For $\text{MAP}_\sigma^{\text{boost}}$, an optimal initialization point $\sigma^0$  was computed by minimizing the error between $\widehat{\hv}$ and a Gaussian of spread $\sigma$.

\section{Numerical  results}

We generated a dataset of 50 $(\xvc,\hvc,\veps)$ triples generating measurements $\yv$. We used images with parameters $n=32$, $K=512$, $b=0.5$ . For the kernels, we used a Gamma distribution with parameters $a=2$ and $\beta=1$. Additive white Gaussian noise with covariance parameter $c_{\veps} = 0.0009$ was considered, which corresponds to a quite high noise level. Such set of simulation parameters were chosen to allow a fair comparison between the MAP and the LMMSE approaches.

A maximum number of $\texttt{max\_iter}=1000$ iterations was performed when running iterative MAP approaches (which led to convergence in almost all cases), alternating 5 steps on $\xv$ with step-size $10^{-1}$ and one step on $\hv$ with step-size $10^{-3}$. These values were chosen to obtain good favorable results for MAP approaches, in comparison to other choices which penalized their quality.
For both $\text{MAP}_\sigma$ and $\text{MAP}_\hv$, the initial guess $\boldsymbol{\alpha^0}$ was sampled from the true prior distribution (still, an ideal choice). Concerning the initialization $\hv^0$ for the kernel, we considered a 2D Gaussian of variance $\frac{a}{\beta}\mathbf{I}$, corresponding to the (ideal) expected value of the corresponding Gamma distribution describing $\sigma$. 

\begin{figure*}[tb]
\center
\includegraphics[width=1\linewidth]{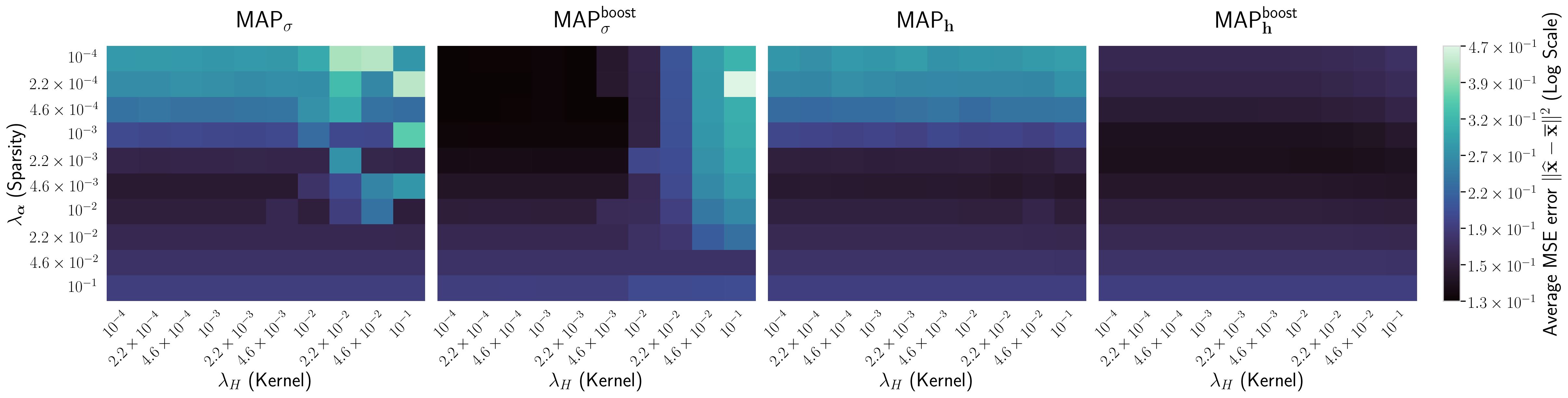}
\caption{Average MSE $\|\hat{\mathbf{x}}-\overline{\mathbf{x}}\|$ on the generated dataset for the four MAP methods considered under various combinations of $\lambda_X$ and $\lambda_H$. LLMSE  initialization (MSE=$0.171$) allows for more stable performance as shown by the flatter error maps. The MSE of the reconstructio obtained by MAP methods in correspondence with the choices of $\lambda_X$ and $\lambda_H$ corresponding to the hyperparameters of the prior functions correspond to a MSE of $0.202$ for  $\text{MAP}_\sigma$. 
}
\label{fig:grid_search}
\end{figure*}

The regularization parameters $\lambda_\alpha$ and $\lambda_H$ were chosen by grid-search over the mean MSE of the reconstructed images of the entire dataset for the four MAP approaches considered, see Figure~\ref{fig:grid_search}. We see that for both $\text{MAP}_\sigma$ and $\text{MAP}_\hv$, there exists an optimal choice for $\lambda_\alpha$ ($10^{-1}$) corresponding to an average error (around 0.155) lower than the LMMSE baseline (0.171). However, most parameters correspond to solutions that are consistently worse than the LLMSE baseline. This highlights the high sensitivity to parameters of MAP approaches overall, net of a limited gain. As a sanity check, we computed also reconstructions in correspondence with the hyperparameters of the underlying distributions ($\lambda_\alpha=\lambda_H=0.0009$), which resulted in an average MSE of $0.202$, which was constantly higher than the one obtained for other parameters. Interestingly, however,  the performance of $\text{MAP}_\sigma^{\text{boost}}$ and $\text{MAP}_\hv^{\text{boost}}$ were more robust w.r.t.~changes in the choice of regularization strengths: even for a suboptimal parameter choice, results were consistently observed to be better than the un-boosted counterpart, with a consistent out-performance of the LMMSE baseline. This test thus suggests that the empirical rule of thumb by which the LMMSE solution could serve as a good initialization point capturing well the low-frequency aspects of both the signal and the kernel, driving the optimization of the non-convex MAP approach towards meaningful local minima with increased high frequencies.

\begin{figure*}[tb]
\center
\includegraphics[width=1\linewidth]{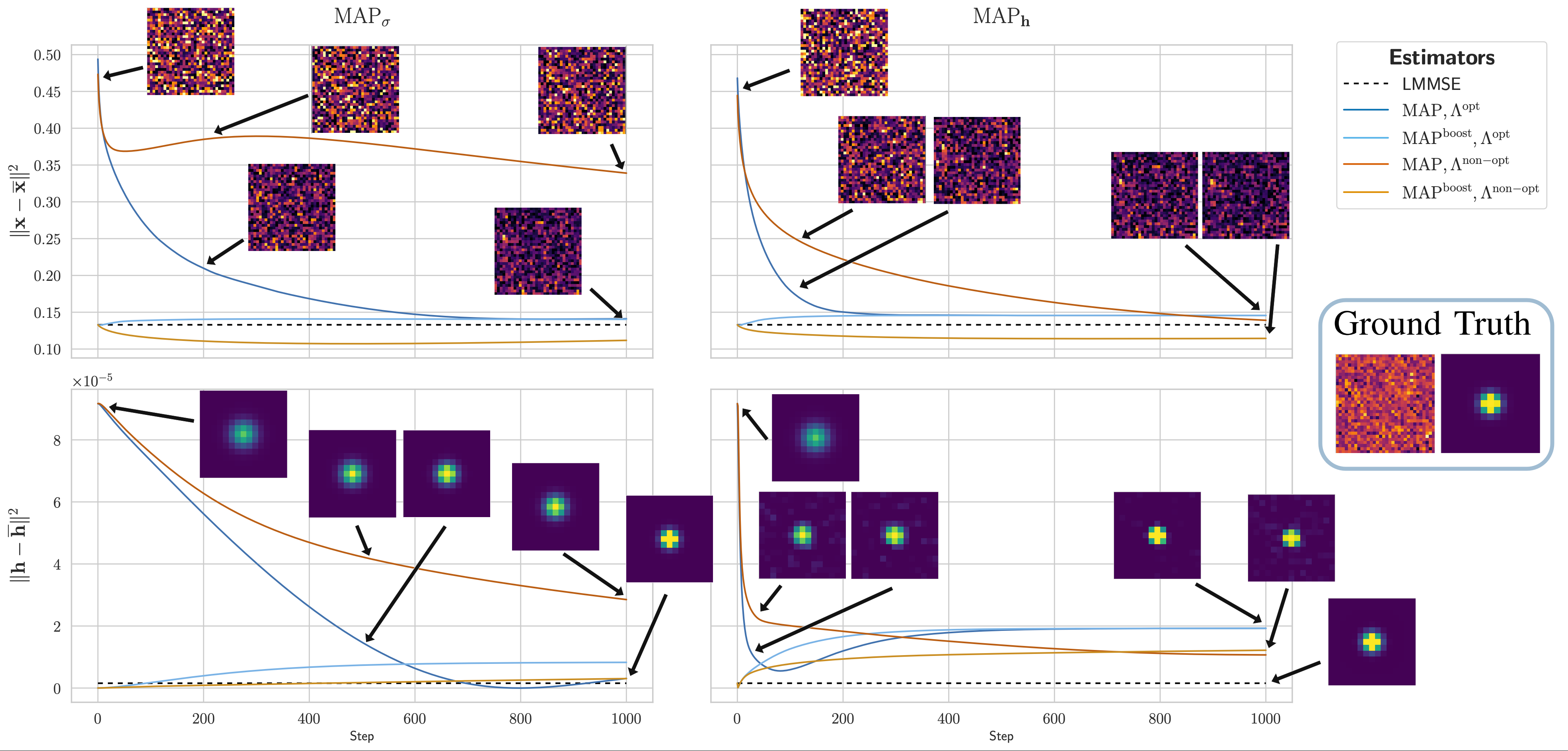}
\caption{Evolution of the MSE values for both the signal and kernel for the $\text{MAP}_\sigma$ and $\text{MAP}_\hv$ approaches and their boosted variants, under different sets of regularization parameters. For the signal, we show the difference image with the ground truth. The LMMSE outperforms MAP approaches, unless regularization parameters are carefully optimized. Using the $\text{MAP}_\sigma^{\text{boost}}$ and $\text{MAP}_\hv^{\text{boost}}$ variants allows for better performances for optimal/non-optimal sets of parameters, thus suggesting that LLMSE solution can be used as a suitable initialization for MAP approaches also when MAP parameters are not properly optimized.}
\label{fig:evolution}
\end{figure*}

 In a different experiment, we compared the evolution of the MSE of both the estimated signal and the estimated kernel along iterations for ``optimal'' regularization parameters (found by grid-search as above) and non-optimal ones for the two approaches $\text{MAP}_\sigma$ and $\text{MAP}_\hv$, with and without the proposed LMMSE-boosting initialization strategy. This gave for $\text{MAP}_\sigma$ and $\text{MAP}_\hv$ two sets of parameters $\Lambda_\sigma^{\text{opt}}= \Lambda_\hv^{\text{opt}}=(0.1,0.001)$ as well as $\Lambda_\sigma^{\text{non-opt}}=(0.0001,0.001)$ and $\Lambda_\hv^{\text{non-opt}}=(0.001,0.001)$, see  Figure~\ref{fig:evolution} for one representative example. 

As a general comment, we notice that when optimized $\text{MAP}_\sigma$ and $\text{MAP}_\hv$ are used to reconstruct the desired signal $\xv$, results will be close to the LMMSE baseline without beating it, while with other parameters they tend to converge to a bad solution. However, using $\text{MAP}_\sigma^{\text{boost}}$ and $\text{MAP}_\hv^{\text{boost}}$ with optimal/non-optimal sets of parameters will either make the algorithm converge to the best MAP solution (but much more quickly), or will evenw improve over this solution. 
Concerning the reconstruction of $\hv$, we noticed that the LMMSE baseline is generally a very good approximation of the unknown, and it does not get improved significantly by running MAP algorithms. 

\begin{figure}[h!]
\center
\includegraphics[width=1\linewidth]{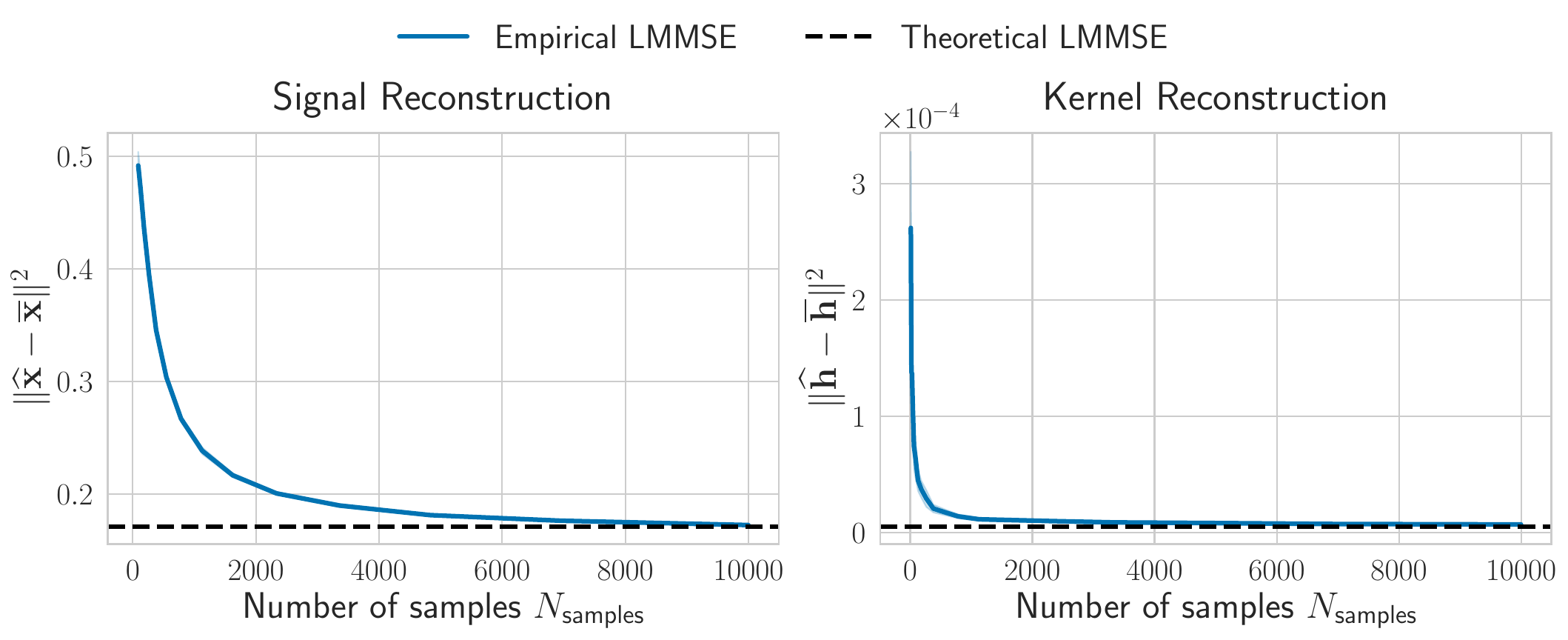}
\caption{MSE of the empirical/theoretical LMMSE constructed for different numbers of samples for the signal (top) and the kernel (bottom). Average of 10 runs. }
\label{fig:lmmse_th_emp}
\end{figure}

Practically, the use of the LLMSE estimator is thus encouraged as a good initialization point. Clearly, whenever no information on the underlying distribution is available, sampled quantities need to be used instead. In this respect, we report in Figure~\ref{fig:lmmse_th_emp} the decay of the empirical LMMSE estimators on both $\xv$ and $\hv$ computed using the empirical first and second moments from a dataset of increasing $N_{\text{samples}}$ examples w.r.t.~the theoretical ones.
As predicted by the theory in \cite{buskulic2025sample}
while the error on $\xv$ decays following an $O(1/N_{\text{samples}})$ decrease, 
the error decreases over $\hv$ in a much steeper way, thus requiring requires less data to achieve almost perfect results. 
The theoretical study of such convergence and its practical use in real-world applications are both interesting questions for future research.

\section{Conclusions}

We validated empirically, in a statistically controlled setting, how MAP-based approach on blind inverse problems are difficult to use and unstable, and this even when the correct regularization functions are used and the setting is chosen to be favorable. On the other hand, we observed how the LMMSE provide a robust estimator for the recovery of $\xvc$ but even more impressively for $\hvc$. We also showed that this estimator can be used as the initialization point of MAP approaches and that it improves significantly their results on a wide variety of regularization parameters. In fact, our best results were obtained using MAP with small regularization as a way to recover high-frequencies and improve consistency with $\yv$. We think the LMMSE offer a promising avenue to stabilize blind inverse problems reconstruction and we would like in the future to better understand its capabilities, especially on the recovery of the kernels, which seems better than on the signals.

\bibliographystyle{IEEEtran}
\bibliography{references}

\end{document}